\documentclass[aps,prl,twocolumn,floatfix,showpacs]{revtex4}
\makeatletter

\usepackage{epsfig}
\usepackage{float}
\usepackage{graphicx}
\usepackage{epstopdf}
\usepackage{epsfig}
\makeatother
\begin{document}
\title{Predicted field-induced hexatic structure in an ionomer membrane}
\author{Elshad Allahyarov}
\affiliation{Department of Physics, Case Western Reserve University, Cleveland,
Ohio 44106, USA, \\
 and Joint Institute for High Temperatures, Russian Academy of
 Sciences, Moscow 127412, Russia} 
\author{Philip L. Taylor}
\affiliation{Department of Physics, Case Western Reserve University, Cleveland,
Ohio 44106, USA}
\begin{abstract}
 
Coarse-grained molecular-dynamics simulations were used to study the
morphological changes induced in a Nafion$^{ \tiny
  \textregistered}$-like ionomer by the imposition of a strong
electric field. We observe  the formation of novel structures aligned along
the direction of the applied field. The polar head groups of the
ionomer side chains aggregate into clusters, which then form rod-like
formations which assemble into a hexatic array aligned with the
direction of the field.  Occasionally these lines of sulfonates and
protons form a helical structure. Upon removal of the electric field,
the hexatic array of rod-like structures persists, and has a lower
calculated free energy than the original isotropic morphology. 
\end{abstract}
\maketitle

The transport of protons through an ionomer membrane is at the heart
of the operation of a polymer-electrolyte fuel cell.   Any procedure
that facilitates the transport would thus be of significant value in
improving the usefulness of this technology.  In this Letter we
describe simulations that indicate that application of a strong
electric field could change the morphology of an ionomer membrane in
such a way as to enhance proton transport appreciably.

The study of proton transport in aqueous ionomers  has received
considerable attention \cite{mauritz2004,kreuer2004,bruijn}.
For perfluoro-sulfonated ionomer membranes such as Nafion$^{ \tiny \textregistered}$
\cite{Gierke1} the structure is characterized by a  long,
hydrophobic, fluorinated  main chain, and short side chains terminating in
hydrophilic sulfonate 
anion groups. The dissimilarity between the hydrophobic backbone and
the hydrophilic side chain terminations creates a  
phase-segregated morphology with sharp interfaces between the
domains. The sulfonic-acid functional groups aggregate to form a hydrophilic
domain that is hydrated upon absorption of water. It is within this
continuous domain that ionic conductivity occurs when protons dissociate
from their anionic counterions and combine with water to become mobile hydronium ions.

Beginning with a seminal paper by Eisenberg \cite{eisenberg}, a number of authors have suggested
morphological geometries for this microphase separation.  These include
inverted globular micelles 
interconnected by a channel structure of cylindrical micelles \cite{Gierke2},  and 
lamellar, sandwich-like structures.  Other recent approaches  are
based on micelle-channel  \cite{ioselevich,Gebel} and inverted micelle
models \cite{shmidt2007}.  There is, however, no complete consensus as
to the structure of the percolating networks of sulfonate or water
channels.   
 
We have performed simulations of almost-dry Nafion$^{ \tiny \textregistered}$-like ionomers
with the goal of determining the morphological changes that would be
produced by strong externally-applied electric fields. We find that  
the side chains  self-organize into cylindrical clusters having their
axes parallel to the applied field, and that these clusters form a
hexatic array 
in the plane 
perpendicular to the field direction. At very strong fields 
each of the rod-like clusters has an ordered inner
core consisting of distinct wires of sulfonate head groups and their
attendant protons. The stable cylindrical cluster structures that
emerged included an unusual 
spiral arrangement of three sulfonic wires
with four protonic chains. The structural changes resulting from the
poling process appear irreversible. 

% figure 1
\begin{figure} [!h]
\hspace{-1.cm}
 \includegraphics*[width=0.43\textwidth]{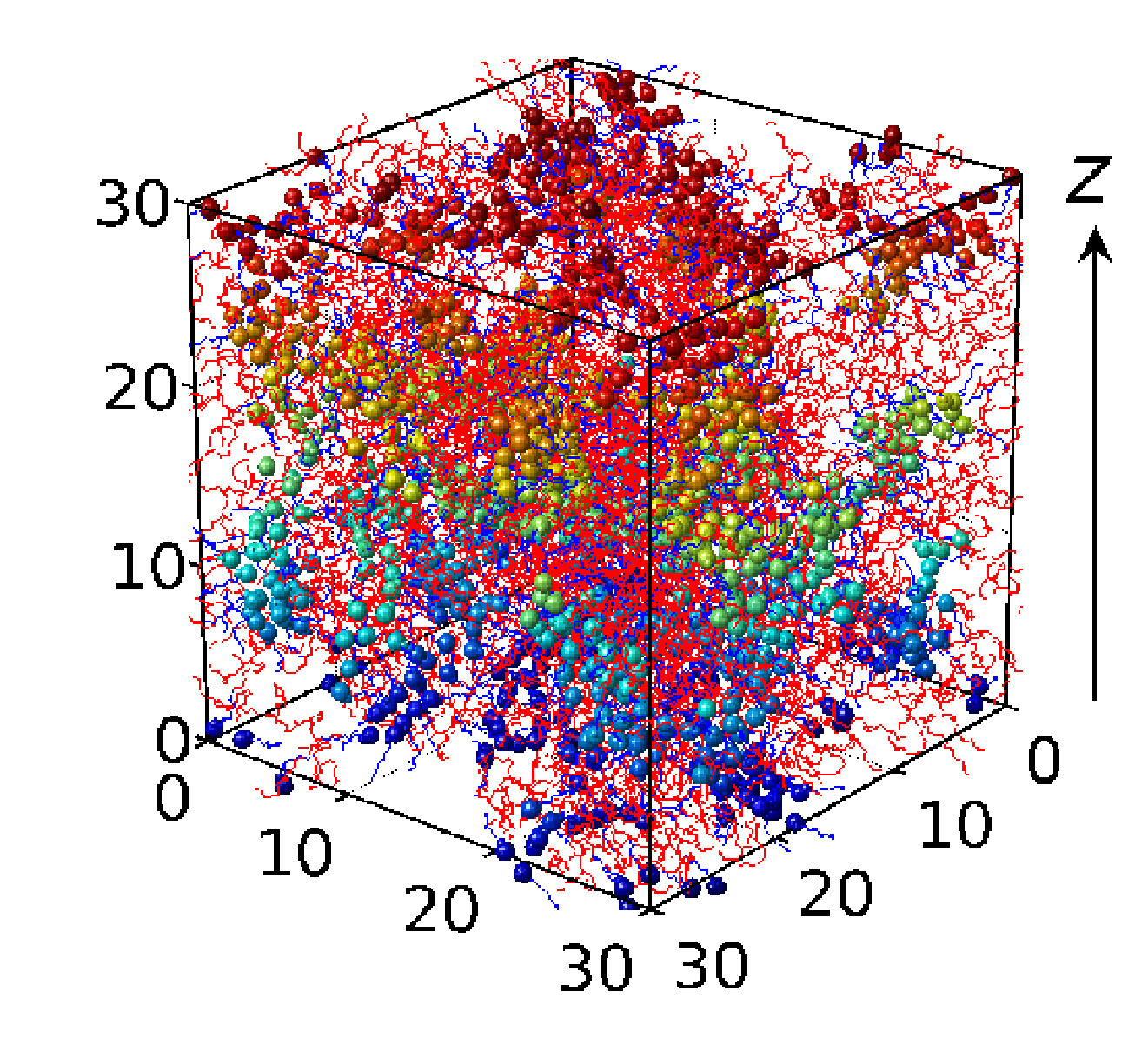}
%{prl_figures/prl_fig1_small.eps}
\vskip 0.5cm \caption{(Color online) This snapshot shows the ionomer
  before application of a field. Large/small spheres represent the
  end-group oxygen/sulfur  
  atoms of the side chains. The colors of the spheres represent
  altitudes, with  blue at the base and red at the top of the
  simulation box.  The size of all structural elements is schematic
  rather than space-filling.  
\label{fig1}}\end{figure}

{\it Model} -- We employ a united-atom representation for Nafion
 \cite{allahyarov-1,wescott2006,vishnyakov2001,yamamoto}, where  
 the CF$_{2}$ and  CF$_{3}$ groups of the backbone and side chains, and
 the sulfur  atom S and oxygen group O$_3$ of sulfonate head
 groups are modeled as Lennard-Jones (LJ) monomers with a universal
 diameter $\sigma=0.35$ nm.   We assume the number $\lambda$ of water
 molecules per sulfonate is small ($\lambda < 5$), and that a proton
 detached from a sulfonate captures a water molecule and becomes a
 hydronium ion.  
%The remaining water is treated as a dielectric
% medium. 
 
The force-field parameters for the ionomer are chosen to agree in most
instances with the Nafion model of Paddison \cite{Paddison1} and are
 given in our recent paper \cite{allahyarov-2}.
  All the partial charges on the
ether oxygen, carbon and fluorine atoms of the side chain, and on the
 fluorocarbon groups of the backbone skeleton  are set
to zero. Electrostatic charge is located
on the sulfur atom (charge  $+e$) and oxygen group (charge 
 $-2e$), such that the total charge of the sulfonate head group  SO$_{3}^{-}$ is $-e$. The 
 hydrogen ion H$^{+}$ is considered as part of a hydronium complex of
 charge $+e$.
% that may be either loosely bound or detached from the sulfonate. 
  For  the low water contents ($\lambda<5$)  considered in this work, 
the remaining water molecules are  mostly  in an immobile state near the pore
walls.  Therefore  one can effectively integrate 
out their degrees of freedom by introducing
a distance-dependent dielectric permittivity \cite{allahyarov-2} and
an effective proton diameter $\sigma$. Aside from these two
approximations, our approach is
 equivalent to the consideration of a membrane with $\lambda=1$, 
when all its sulfonic groups are dissociated forming hydronium ions
\cite{dokmais-2007,laporta-1999}.

% figure 2
\vskip-0.3cm 
\begin{figure} [!h]
\includegraphics*[width=0.25\textwidth]{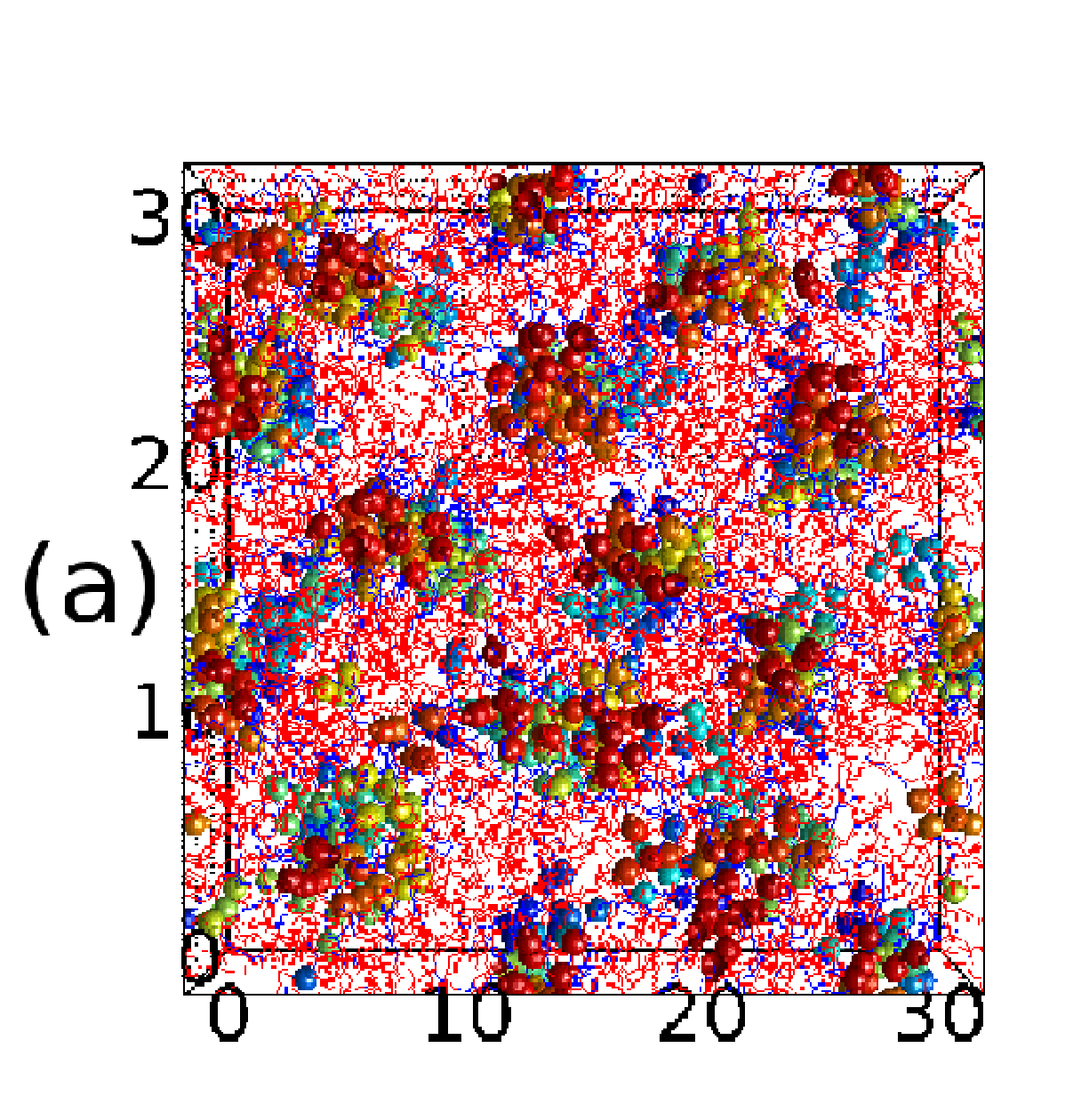}\includegraphics*[width=0.25\textwidth]{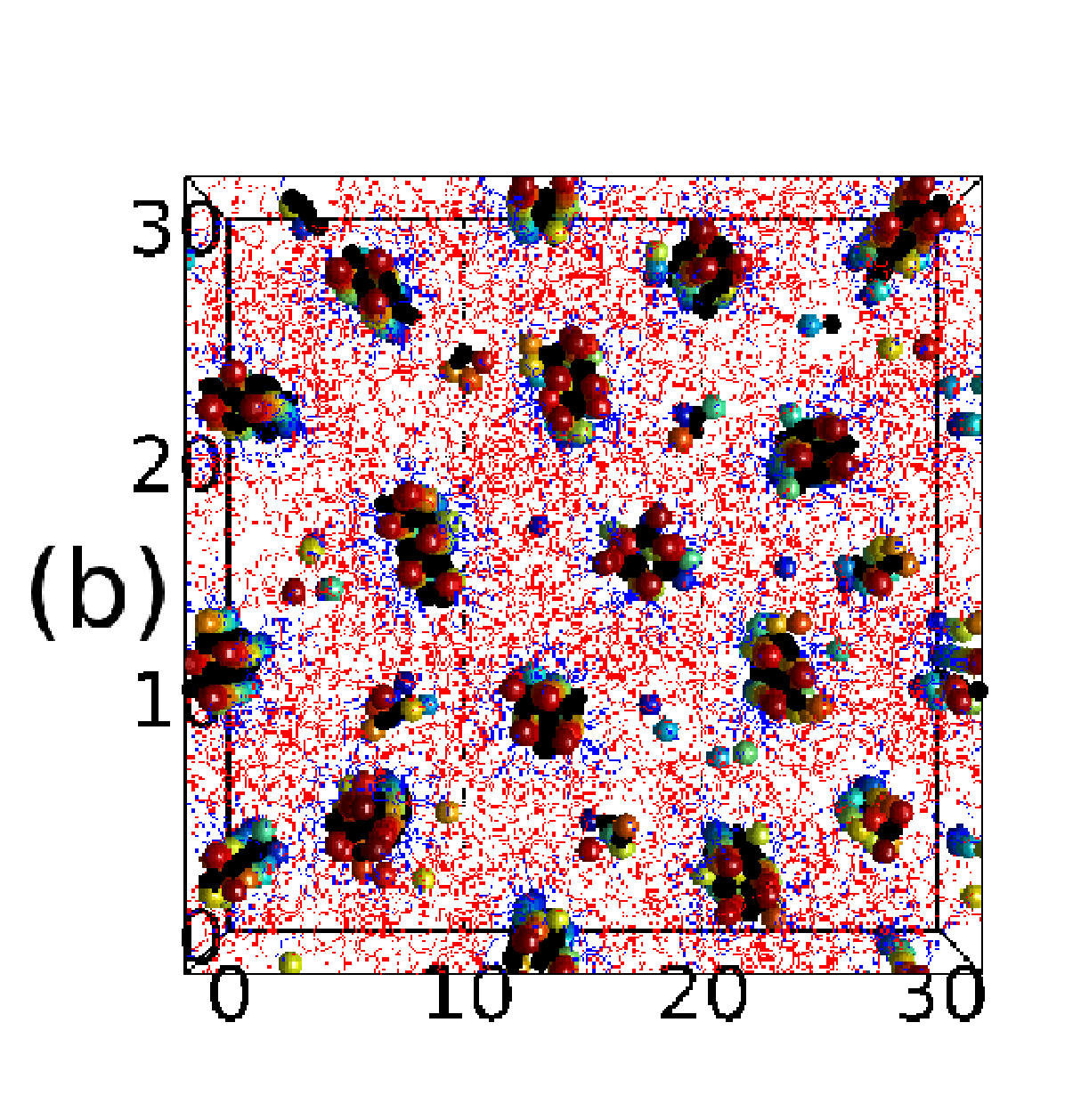}   
%{prl_figures/prl_fig2_small.eps}{prl_figures/prl_fig2b_small.eps}  
\vskip-0.5cm \caption{(Color online)   View from above of 
 the simulation system for applied fields of (a)
  $1.2\times 10^{6}$ kV/m  and (b) $E=8.2\times 10^{6}$ kV/m. Again,
 a blue color represents a low altitude and a red color a high
 altitude.  Protons  are shown as 
  black dots in diagram (b).
\label{fig2}}
\end{figure}

 The total potential energy of the membrane is taken to  be  
\begin{equation}
U(\vec r) = \displaystyle\sum_{i}{U_b^i} +
\displaystyle\sum_{j}{U_{\theta}^j} + 
\displaystyle\sum_{m}{U_{\varphi}^m} + 
\displaystyle\sum_{k,l}{U_{nb}{\left( |\vec r_k -\vec r_l| \right)}} 
\label{eq1} 
\end{equation}
where $(\vec r_1, \vec r_2,...,\vec r_N)$ are the
three-dimensional position vectors of the $N$ particles in the system.   In Eq.~(\ref{eq1}),
$i$ runs over all bonds, $j$ runs over all bond angles, $m$ runs over all
torsional angles, and $k$, $l$ run over all non-bonded (Lennard-Jones
and Coulomb) force center pairs in the system.
For lightly humidified membrane materials,  the protons
are mostly energetically bound  to their host sulfonate groups
\cite{jalani2005,thompson2006}.  
The resulting SO$_3^-$H effective dipoles have  zero net charge,
and thus experience no net force in a uniform applied external field,
but are subject to torques. 
Interactions between dipoles are screened with a  dipolar screening length
 $R_s$ \cite{jonscher2001}.  

In  direct-current experiments, the electrostatic potential
$U_{\rm ext}({\vec  r})$ inside the polymer is 
considered to be linear \cite{enikov}. In  non-equilibrium molecular
dynamic computer simulations the
current flow is usually induced by 
applying an external electrostatic field $E$ \cite{franco2006}.    
The scalar protonic conductivity $\chi$ is defined from 
the linear relation $ \vec j = \chi \vec E$  between the applied field
$\vec E$ and the induced current density $\vec j$ in the membrane. When the external field is
applied along the {\it z}-axis,  $j =  \sum q_i v_{i,z}/V$, 
 where $v_{i,z}$ is the axial component of the velocity of the $i$-th
 proton, $i=1,..,N_s$, with $N_s$ the total number of protons and $V$  the sample volume. 
The conductivity can also be related to the diffusivity $D$ through the
Nernst-Einstein relation $\chi$=$N_S e^2 D/V k_BT$
in equilibrium molecular dynamics simulations. 
However, this method is less accurate for phase
 separated systems, where confinement effects, such as proton
 trapping in the water-sulfonate clusters, strongly affects the overall
 diffusion constant $D$. 
% figure 3
\vskip-0.4cm 
\begin{figure} [!h]
\includegraphics*[width=0.245\textwidth]{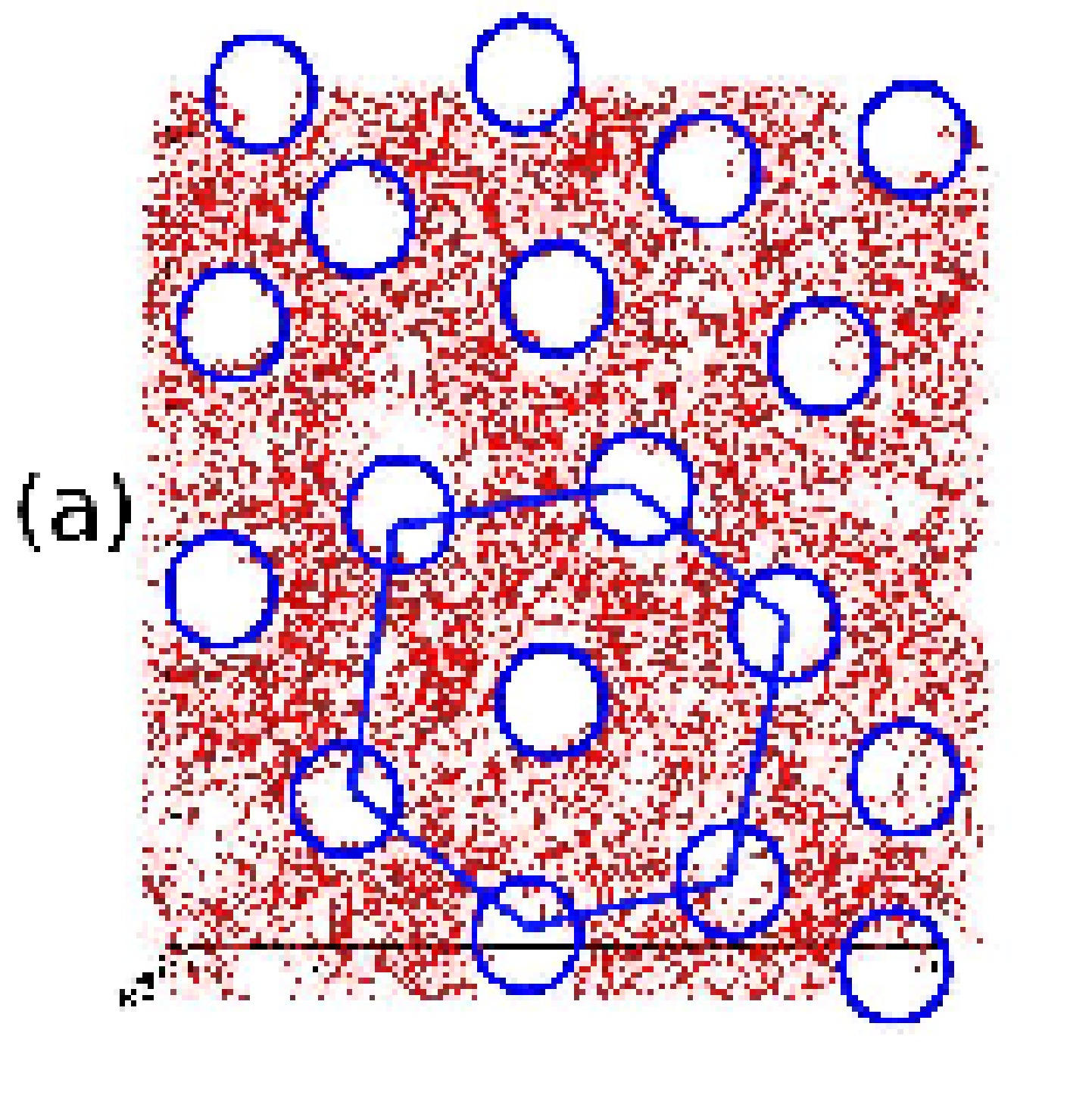}\includegraphics*[width=0.25\textwidth]{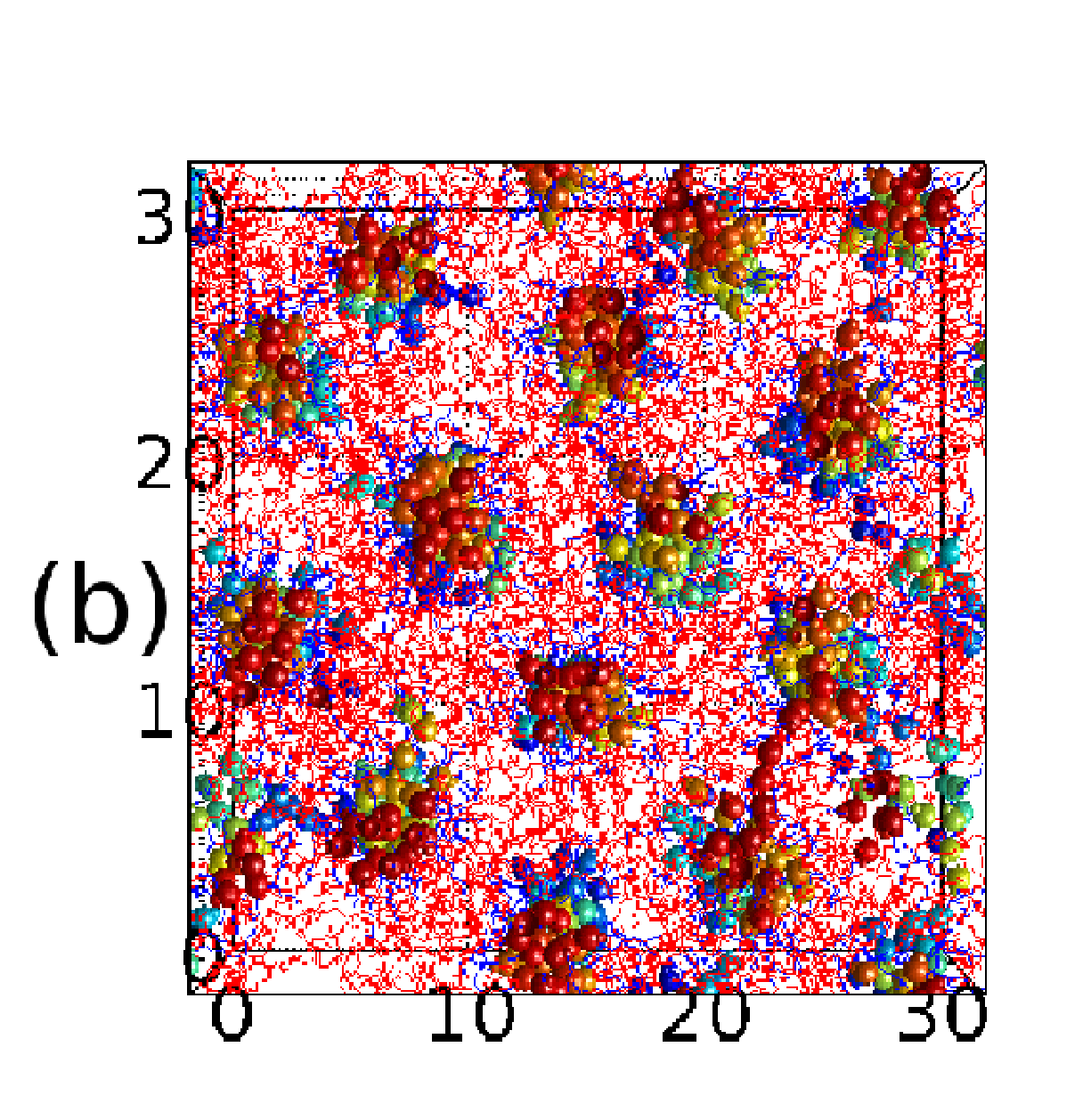} 
%{prl_figures/prl_fig3a_small.eps}{prl_figures/prl_fig3b_small.eps} 
\vskip-0.5cm \caption{(Color online)  (a) Depiction of only the
  backbone polymer shows the hexatic array of proton channels formed
  by the sulfonates.    Circles and solid lines are guides for 
the  eye. (b) Remanent structure in a membrane after removal of the
  applied field. \label{fig3}}  
\end{figure}
\vskip-0.2cm

 At low $\lambda$ we expect surface hopping to be
 the leading contribution to the current through the membrane. In this
 case the protons move in the electrostatic 
energy landscape of the sulfonate ions.  Proton transfer is
 expected to occur at  sulfonate-sulfonate separations 0.7-0.8 nm
 \cite{tanimure2004} with an energy  barrier less than 2.1 kcal/mol
 \cite{thompson2006,tanimure2004,commer2002,spohr2006}. 
 We neglect the contributions to the current arising from Grotthuss and en-masse proton
diffusion, which are only significant in highly humidified membranes ($\lambda
\ge 5$), and whose evaluation requires full simulations with explicit water molecules. 
 % figure 4
\begin{figure} [!h]
\includegraphics*[width=0.23\textwidth]{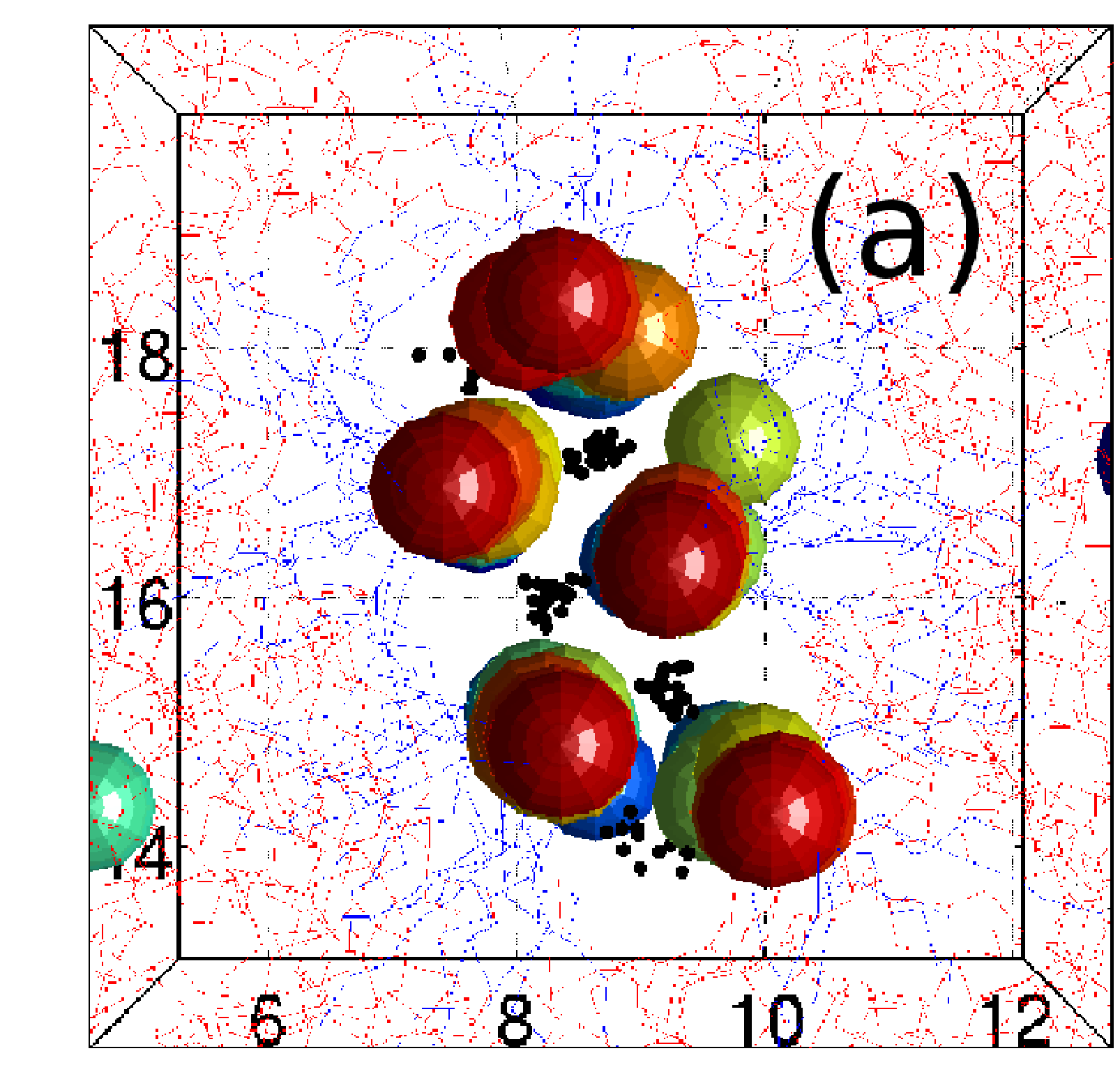}\includegraphics*[width=0.23\textwidth]{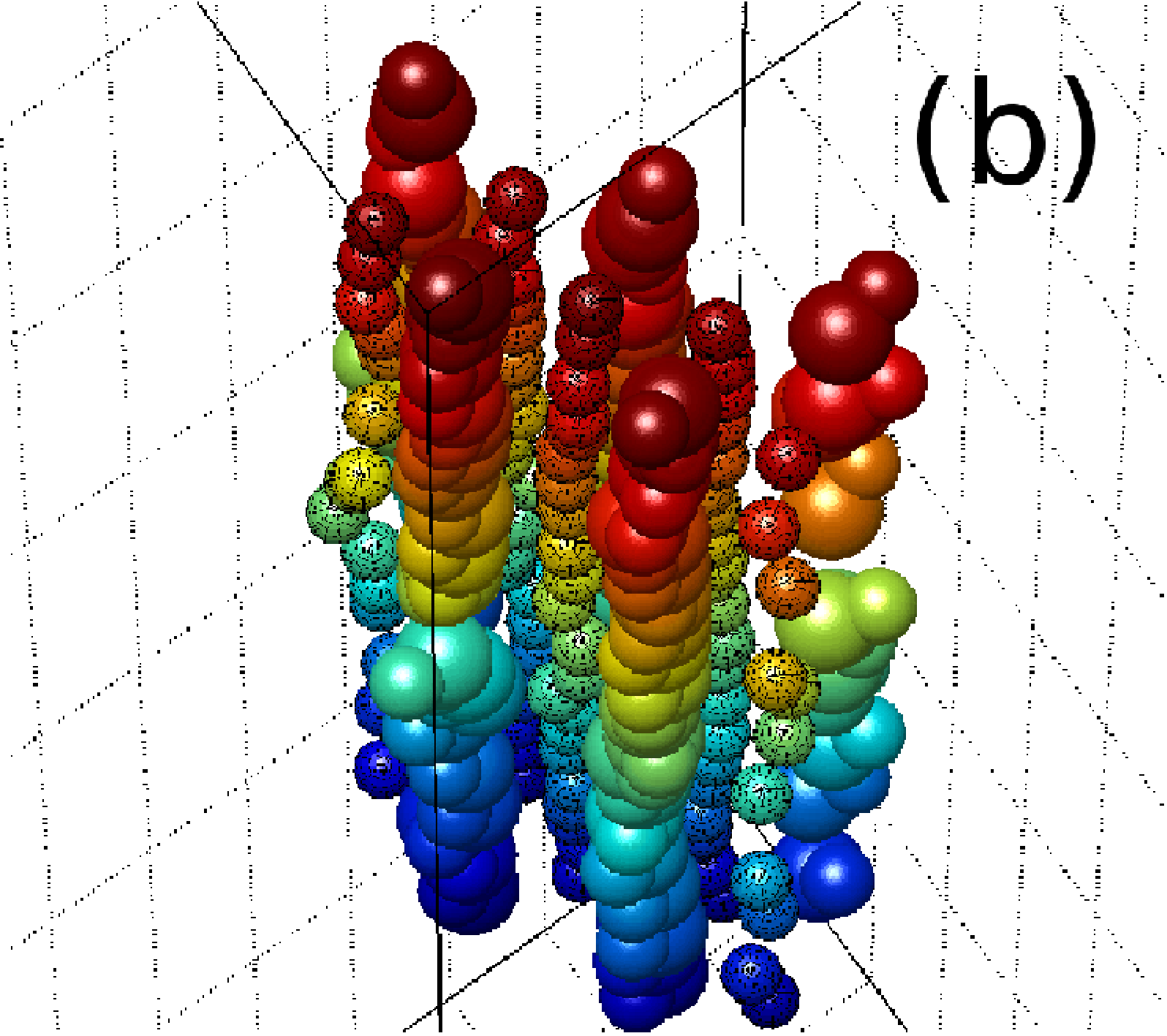}
%{prl_figures/run6-five_small.eps}{prl_figures/five_wires_small.eps}
\vskip-0.5cm \caption{(Color online) The {\it xy} plane projection (a) and 
a perspective view (b) of a single rod-like cluster  with a five-wire
structure from Fig.~\ref{fig2}(b). Small
spheres attached to oxygens represent sulfurs and netted small balls
are protons in the right figure. \label{fig-five}} 
\end{figure}
%\clearpage
% figure 5
\begin{figure} [!h]
\includegraphics*[width=0.23\textwidth]{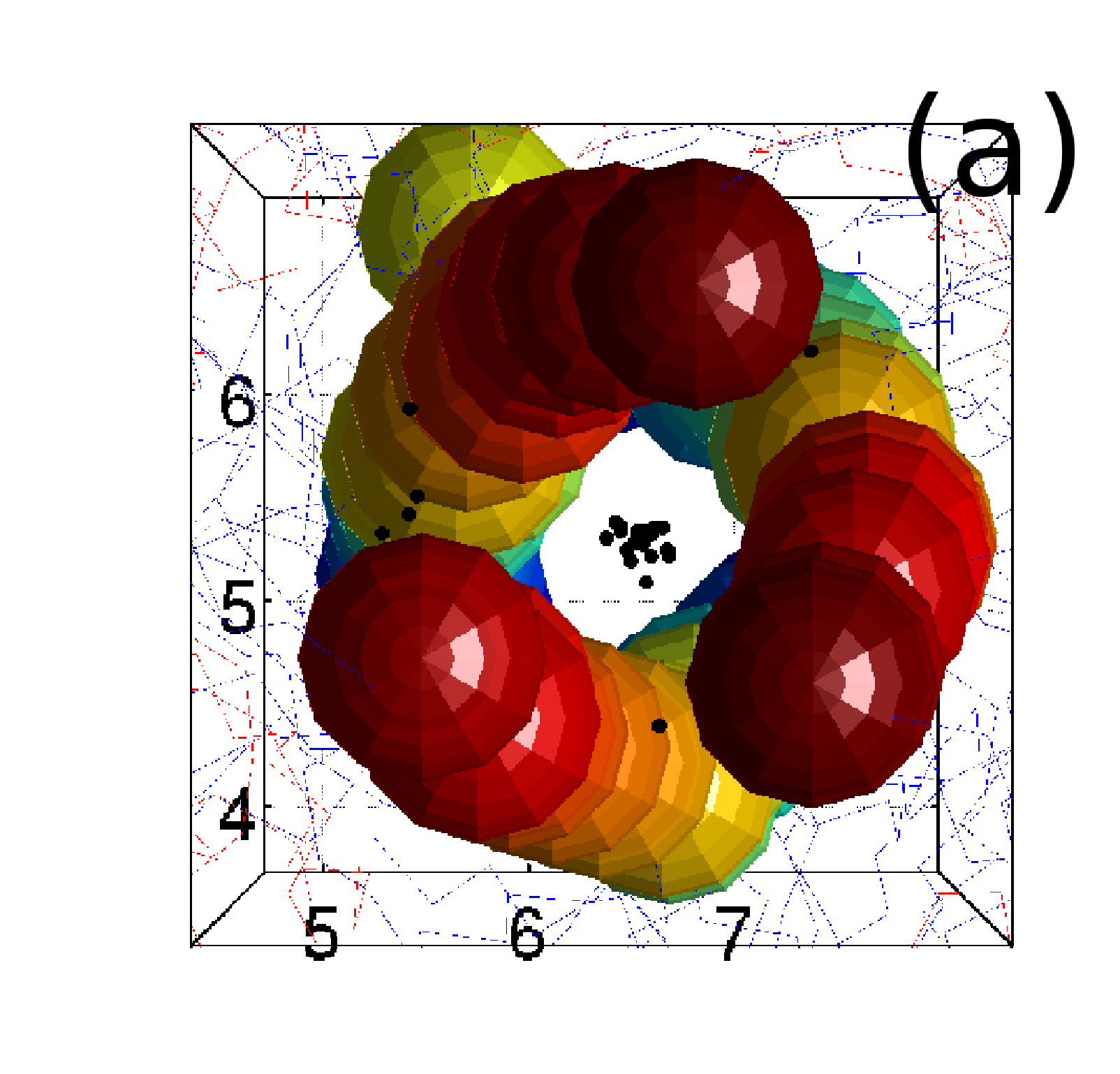}\includegraphics*[width=0.23\textwidth]{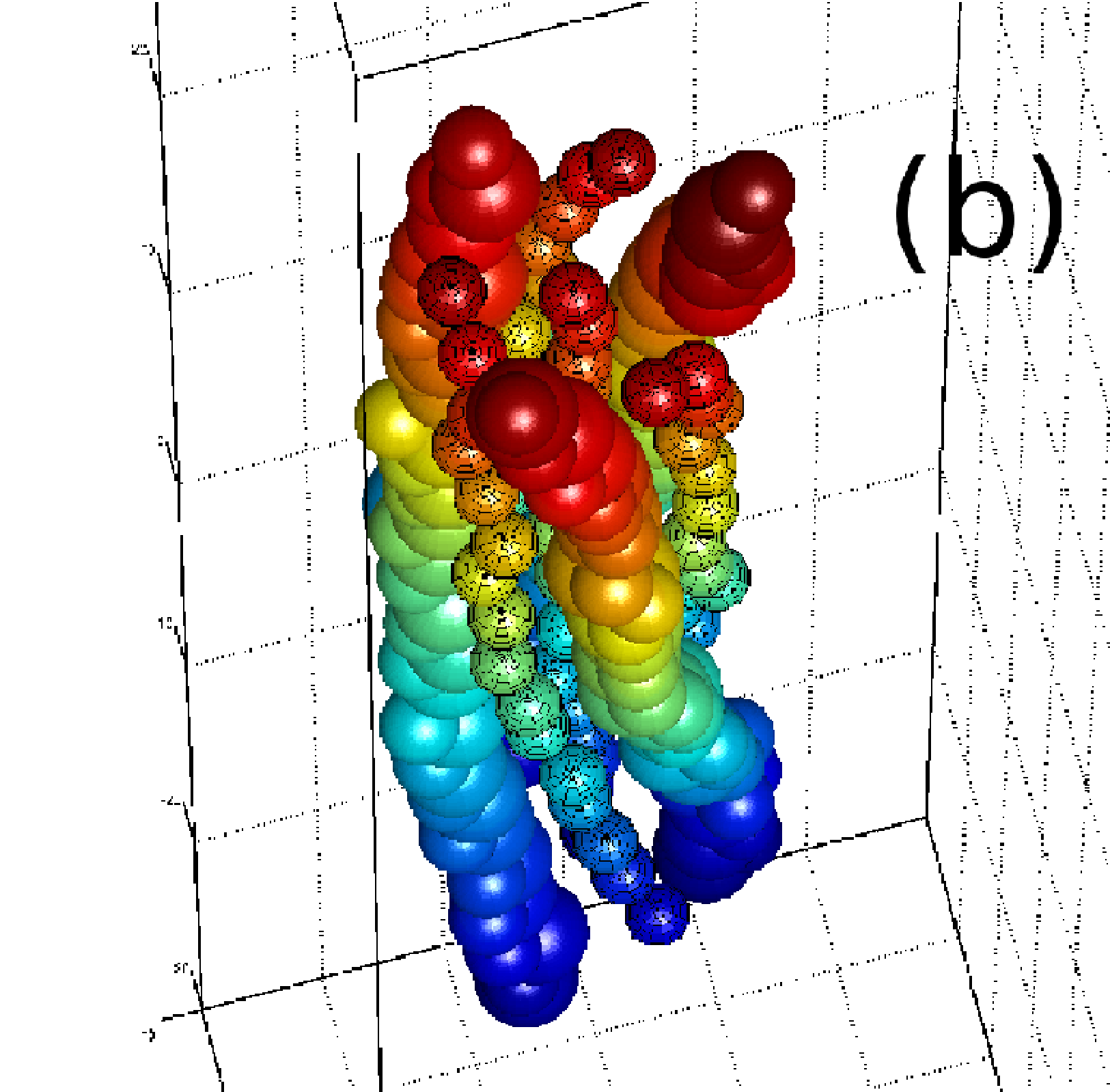}
%{prl_figures/run6-three_small.eps}{prl_figures/spiral_wires_small.eps}
\vskip-0.5cm \caption{(Color online)  The {\it xy} plane projection (a) and 
a perspective view (b) of a single rod-like cluster  with 
a  spiral three-wire structure from Fig.~\ref{fig2}(b). 
 Note that there are four  protonic wires for a triplet of sulfonic
 wires. \label{fig-three}}\end{figure} 
%\clearpage
% figure 6
\begin{figure} [!h]
\includegraphics*[width=0.4\textwidth]{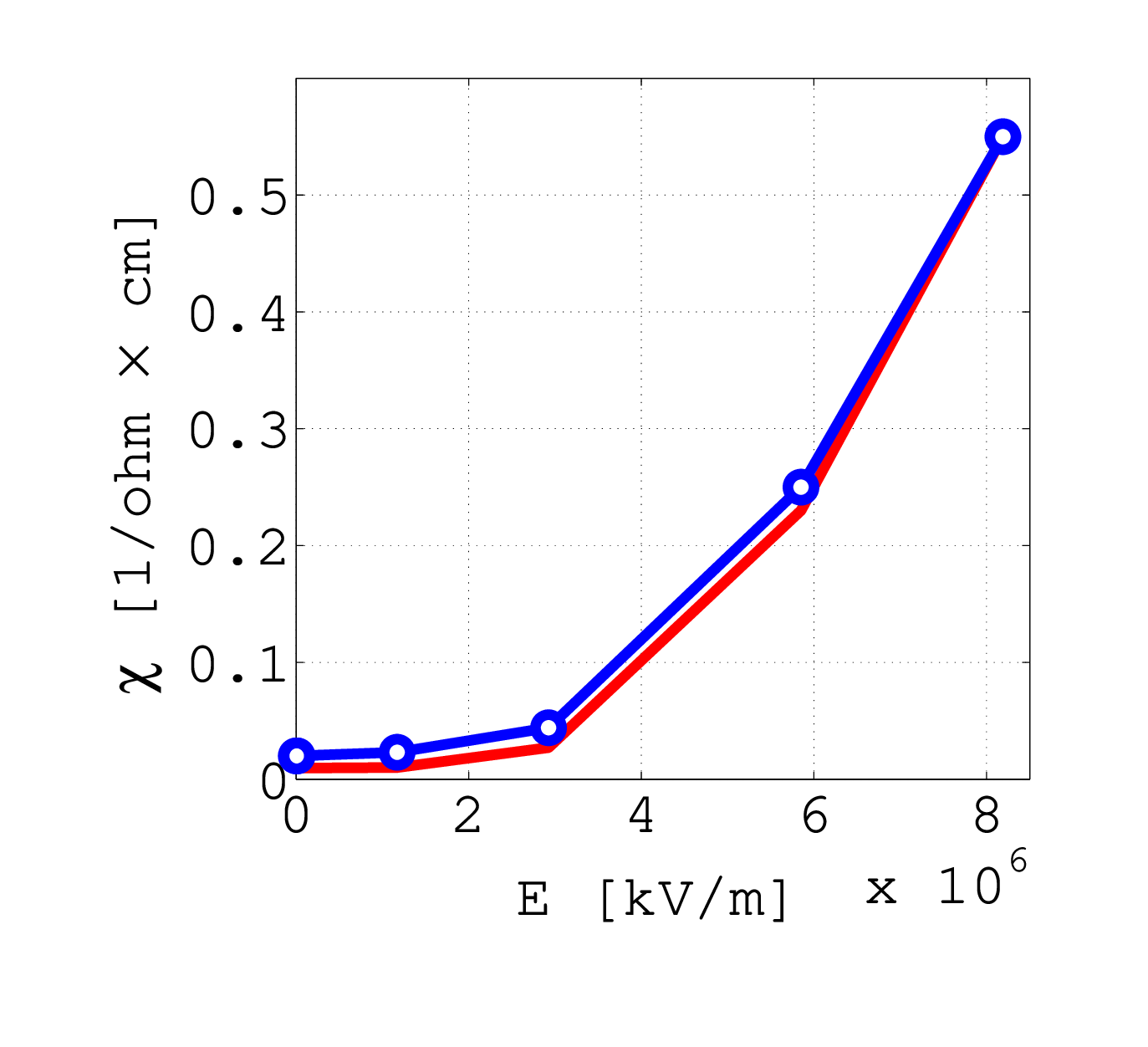}
%{prl_figures/conductivity.eps}
\vskip-0.5cm \caption{(Color online) Protonic conductivity $\chi$ of
  the membrane as a function of 
  applied field $E$. Line without/with circles depicts 
  increasing/decreasing field. \label{fig-current}}
\end{figure}
% figure 7
\begin{figure}
\includegraphics*[width=0.25\textwidth]{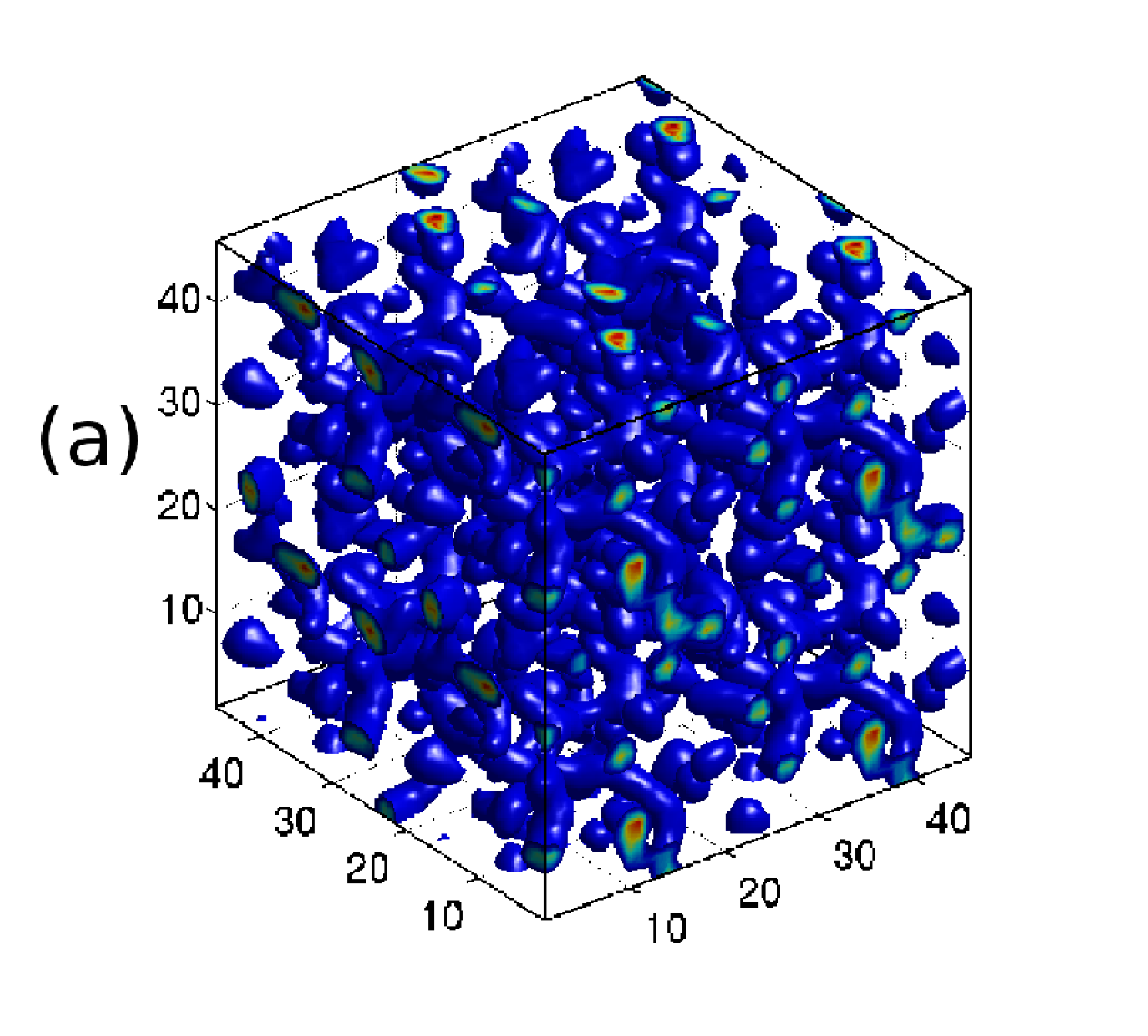}\includegraphics*[width=0.25\textwidth]{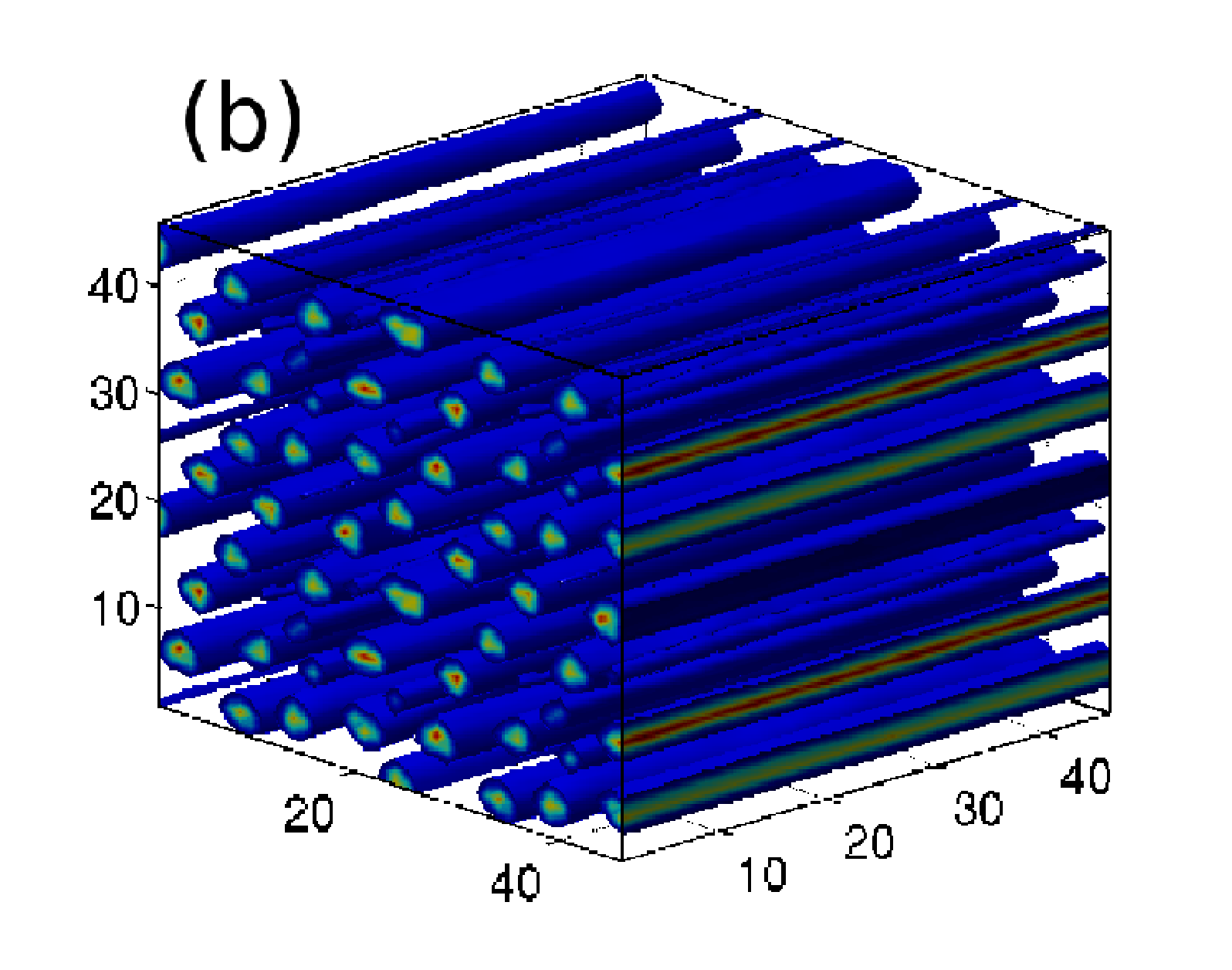} 
%{proton_3D_E_0001_small.eps}{proton_3D_E_07_small.eps} 
\vskip-0.5cm \caption{(Color online) A short-time average of the 3D
  density distribution 
  of protons for material in zero field (a) and  with $E=8.2\times 10^{6}$ kV/m (b).
  \label{fig-3D-proton}}
\end{figure}
%\vskip-0.5cm

{\it{Simulation results}} -- Molecular dynamic 
simulations were performed for coarse-grained membranes kept at constant
volume $V=$(11 nm)$^3$ and constant temperature $T$=300 K. 
 Each side chain contains 2 hydrophilic and 7 hydrophobic monomers, and 
there were 14 hydrophobic backbone monomers between adjacent
sidechains. The system temperature was controlled 
by a Langevin thermostat with a friction coefficient
$\gamma=0.1$ and Gaussian noise $6k_{B}T\gamma$.
The equations of motion were integrated using the velocity Verlet
algorithm with a time step of 0.2 fs. Periodic
boundary conditions and Lekner summation \cite{Lekner} of
long-range electrostatic interactions were used.

 We temporarily detached the side chains from
the backbone skeleton \cite{rivin-vishnyakov} and  cut the backbone into 14-monomer
segments \cite{glotzer2002} for initial equilibrating runs of 50 ps,
after which the polymer was reassembled and 
 equilibrated with another
50 ps  run.   Statistically averaged quantities were gathered
during the next 5-10 ns.
A snapshot of a simulated system in the absence of an applied field is given in
Fig.~\ref{fig1}. The ionomer
has undergone a partial phase separation on the nanometer scale
into a polymer phase consisting of  backbone with pendant side chains
and a hydrophilic phase formed by loosely connected clusters of the
sulfonate head groups, which  are mostly composed  
of compact multiplets of about 12 sulfonates.  

Application of a strong electric field  of the order of $10^9$ V/m gives rise to 
 the formation of the new morphology illustrated in
Fig.~\ref{fig2}. The isotropic system of clusters 
 in Fig.~\ref{fig1} is now replaced by a system of cylinders aligned
 along the {\it z}-axis. 
The sulfonates form the walls of these cylinders, which have
a diameter of $\approx$5$\sigma$ ($\approx 1.8$ nm) and a separation
 of roughly 8-10$\sigma$.   As the field strength is increased, these
 cylinders form a hexatic array, as shown in   
Fig.~\ref{fig3}(a).   At the same time, the internal structure of the
 cylinders self-organizes into an array 
 of distinctive `wire-like'
chains of sulfonates, or to be more precise, chains of oxygens and
 sulfurs, and parallel chains of 
 protons.  
 %  What does this mean?   Also, two adjacent sulfonic wires together with the
% protonic wires assigned to them form a simple cubic structure. 
 Though the number of sulfonic wires 
per cylinder varies between 1 and 5, nearly half of the 
structures have a  three-wire structure. 
 All of the other structures observed 
  consist of an equal number of
 sulfonate and proton wires, an example for a five-wire structure
 being given in Fig.~\ref{fig-five}.  The mean SO$_3^-$--SO$_3^-$
 group spacing along the chains is 
about 0.65-0.80 nm, which is close to the separation distance expected for
Nafion-like membranes \cite{tanimure2004}. 
 The cylinders composed of three sulfonic wires accommodate four protonic wires, as shown in
 Fig.~\ref{fig-three} for the special case of a spiral
 cylindrical cluster.  The apparent stability of this structure may be
 an artifact of the  periodic boundary conditions.
 
A remarkable result of the simulations was that this field-induced
 morphology remains after the removal of the applied field.
 Fig.~\ref{fig3}(b) shows the ordered structure that persists after
 the field has been removed from the system shown in
 Fig.~\ref{fig2}(b).   In order to understand this  phenomenon we 
 calculated and compared the free energies $A$ of the original
 isotropic structure and the structure with remanent order. 
 The method of thermodynamic integration tells us that the difference
$\Delta A$ between a system with and without Coulomb interactions is
 equal to $\int_{\xi=0}^{\xi=1} \left <  \partial 
H(\xi)/\partial \xi   \right > d \xi$ when the coupling parameter
 $\xi$  characterises the strength of  
  Coulomb interactions $U_c^{ij}$ between head groups $i$ and $j$. The
  configurational part of a parametrized
  Hamiltonian $H(\xi)$ was chosen as $U(\xi,\vec r) = K + U(\vec r) + \left ( \xi-1
  \right ) \sum_{i>j}{U_c^{ij}}$.   
 The calculated values of $\Delta A/Nk_BT=-9.3$ for
the untreated membrane and  $\Delta A/Nk_BT=-9.7$ for the treated membrane
indicate that the latter has a lower free energy.
 The morphological changes induced by the strong external field thus appear to be irreversible.

%evaluating the average effective acceptance ratio over
%the Boltzmann distribution function, $A/Nk_BT=\ln \langle \exp \left( U - \bar{U}\right)
%\rangle + \bar{U}$. Here $A$ is the Helmholtz free
%energy, $U$ and $\bar{U}$ are the instant and average potential
%energies of the membrane. Calculated values of $A/Nk_BT$=-4.28 for
%standard membrane and  $A/Nk_BT$=-4.41 for the pre-exposed membrane
%indicate that the latter has energetically favorable configuration. 
%Thus, the morphological changes induced by external fields
%are shown to be `{\it irreversible}'.

Proton conduction is strongly enhanced by the formation of the
sulfonate cylinders along the current direction, as indicated in
Fig.~\ref{fig-current}, which shows that  the effective conductivity
$\chi(E)$ increases rapidly with field strength.  Because the effect
of high fields is irreversible, there is hysteresis in the
current-field plot. 
Both the small-field conductivity $\chi=9.5\times10^{-3}$ S/cm of the
isotropic membrane and the value of 
$\chi=2\times10^{-2}$ S/cm for the poled membrane are much
 smaller than the reported experimental values $\chi=0.1$ S/cm for humidified
 Nafion, but much larger than the theoretically predicted surface
 hopping conductivity $\chi \sim 10^{-5}$ S/cm \cite{chan2004}.

The 3D density distribution of protons, plotted in
Fig.~\ref{fig-3D-proton}, shows a  disordered
distribution of protonic clouds inside the isotropic material.  In the
presence of a strong applied field, however,  the 
protonic clusters connect to each other and form large clusters. These
clusters elongate along  the field direction, 
eventually forming  rod-like channels.

We thus conclude from our simulations that application of a strong
applied electric field to a Nafion$^{ \tiny \textregistered}$-like
ionomer may induce the formation of rod-like structures of sulfonate
groups,  which assemble into a hexatic array aligned with the
direction of the field.  This novel morphology persists after the removal of
the field as a consequence of the lowering of the free energy
associated with this change in structure. 
The induced agglomeration of
hydrophilic head groups into long rods appears to reduce  the percolation
 threshold for ion conductance \cite{oren}. 
 Simulations also indicate that strong
correlations between the sulfonate head groups and protons inside
 the rods create wire-like structures of charged particles. The geometry of
these structures is very rich and includes stable spiral-like three-wire 
clusters. 
 These effects, if confirmed experimentally, may have an impact on the
 industrial applications of Nafion-like 
membranes. 
      
This work was supported by the US Department of Energy under Grant
DE-FG02-05ER46244, and made  use of facilities at the
CWRU ITS High Performance Computing Cluster and the Ohio Supercomputing
Center. Discussions with G. Buxton, E. Spohr, P. Pintauro,
R. Wycisk and M. Litt are gratefully acknowledged. 
%\clearpage
%\pagebreak[4]
%\vskip-0.5cm

\end{document}